\documentclass[%
 reprint,
 amsmath,amssymb,
 aps,
]{revtex4-1}

\usepackage{graphicx}
\usepackage{dcolumn}
\usepackage{bm}

\begin{document}

\preprint{APS/123-QED}

\title{Quantitative understanding of the onset of dense granular flows}

\author{Kasra Farain and Daniel Bonn}

\affiliation{Van der Waals–Zeeman Institute, Institute of Physics, University of Amsterdam, Science Park 904,  1098XH Amsterdam, The Netherlands\\
 k.farain@uva.nl\\d.bonn@uva.nl}

\begin{abstract}
The question when and how dense granular materials start to flow under stress, despite many industrial and geophysical applications, remains largely unresolved. We develop and test a simple equation for the onset of quasi-static flows of granular materials which is based on the frictional aging of the granular packing. The result is a non-monotonic stress-strain relation which – akin to classical friction – is independent of the shear rate. This relation suffices to understand the below-threshold deformations of aging granular media, and its solid-to-liquid transition. Our results also elucidate the (flow) history dependence of the mechanical properties, and the sensitivity to initial preparation of granular media.
\end{abstract}

\pacs{Valid PACS appear here}
\maketitle



Under small stresses, the physical state of an accumulation of solid grains, sand for example, can evolve or age slowly over long periods of time, in a phenomenologically similar way as an ensemble of atoms or molecules in electron glasses \cite{1} and polymers \cite{2}. Such slow relaxation and aging phenomena in granular media, as non-equilibrium systems of discrete grains, have been of considerable interest in the past years to understand the underlying statistical principles of glassy and jammed dynamics generically  \cite{3, 4, 5, 6}. The evolution of granular systems, on the other hand, are critically important in their own right due to many industrial and geophysical applications. The problems of stability and flow of granular matter arise in systems as diverse as cosmetics and foods, nuclear reactors  \cite{7}, soil mechanics  \cite{8,9}, and earthquakes  \cite{10, 11, 12}. A well-known example is the settling of soil; newly created Dutch polders, made by projecting sand onto an area, are left to settle for several months or more before building can begin; otherwise they are too soft.

Granular materials are often thought of as yield stress materials. They can behave solid-like under small forces, e.g. forming granular piles with a finite slope. But, they can also flow more or less like a liquid when subjected to a shear stress above some threshold value, e.g. in an hourglass. The solid-like phase resists deformation due to the presence of a force-bearing network of grains that are in frictional contact with each other. The strength of the system is, therefore, controlled by the stresses that these frictional contacts can withstand. When enough of these contact points fail under large stresses, the whole microstructure fails in an avalanche like behavior and the material yields to flow  \cite{3, 13}. The rheological properties of such ‘rapid’ granular flows, which are characterized by inertially-dominated and hence rate-dependent dynamics, have been investigated extensively in different settings; and constitutive equations, based on dimensional analysis, have been proposed for the relation between stresses and shear rates \cite{13, 14, 15, 16, 17}.  The quasi-static deformations of the solid-like phase close to but below the threshold, and the solid to liquid transition itself, however, despite much work  \cite{18, 19, 20, 21, 22}, still lack a unified, empirically supported, theoretical framework. 

The onset of granular flow is often described by a friction coefficient \cite{15, 23, 24}, the ratio of shear to normal stresses at the failure point or, equivalently, the tangent of the angle of repose of the material. Such zero-order description, although it has had success in describing the aging and humidity-dependence of the granular friction \cite{23, 24}, suggests that there is no flow before the complete failure of the whole network. Yet, granular media can move slowly or creep even under stresses below the yield strength  \cite{18, 25, 26, 27}. Such below-threshold deformations are evidently rate-independent, contrasting with the inertial rapid-flow regime \cite{18, 28, 29}.

Here, we exploit the similarities between granular materials and a larger subclass of materials, thixotropic yield stress materials, with a time-dependent structure that couples with the flow \cite{30}, to develop a simple analytical relation for the onset of quasi-static flows of granular materials. The interactions between the grains in the slow flows considered here occur solely through frictional sliding (not collision); and the inertia can be safely ignored in such flows. The relation predicts a stress overshoot for the onset of dry granular flows, qualitatively similar to those in the yield stress fluids and dense suspensions \cite{31} but with a shear-rate scaling that originates from the dry friction between the grains. We subsequently test our model by comparing it to slow shear experiments on a granular medium of hard spherical particles. We find that the model quantitatively explains the stress response of the granular matter subjected to small shear rates. The model also accounts for the microscale below-threshold deformations and explains the threshold behavior. 

The model. — We begin our formulation of the onset of quasi-static flow by specifying the basic assumptions that are considered in the model. First, it is assumed that the granular system experiences a logarithmic aging under stresses below the threshold (‘yield’) stress \cite{23, 32, 33}. This may be written as
\begin{equation}
	F_{v=0}=a~\ln(\frac{t}{\tau}),
\end{equation}
where $t$ is the time since the preparation of the system in a certain state, $\tau$  is a reference time scale that makes $t$ dimensionless and $a$ is a constant. By taking the derivative of this equation, the part of the friction evolution that is imposed by the inherent aging is $dF/dt=a/t$. This means that granular friction starts with a dramatic increase which slows down quickly. The second assumption is that, just like the case of the thixotropic yield stress fluids \cite{30}, the effect of shear is a continuous ‘rejuvenation’ or de-aging which is proportional to the sliding velocity, $v$. Finally, the rejuvenation rate toward the steady-state flow is taken to be also proportional to how far the current state of the granular system is from the steady-state flow, i.e. to $F(t)-F_{d}$, where $F_{d}$ is the steady-state frictional resistance of the system. Therefore, aging with a rate of $a/t$ strengthen the granular medium, while shear rejuvenates it with $\alpha v (F-F_{d})$, where $\alpha$ is a shear constant of the granular material with dimension of $m^{-1}$. This leads to a differential equation for the friction force:
\begin{equation}
	\frac{dF}{dt}=\frac{a}{t}-\alpha v (F-F_{d}),
\end{equation}
that is similar (but not identical) to the one proposed for thixotropic yield stress fluids \cite{30}. In our granular case, solutions are of the form 
\begin{equation}
	F(t)=a e^{-\alpha v t} \int_{t_{0}}^t \frac{e^{\alpha v t'}}{t'}~dt' +F_{d}(1-e^{-\alpha v t}),
\end{equation}
where $t_{0}$ is a constant that determines the initial state of the granular system. This constant can be considered as equivalent to the initial state variable in the rate-and-state equations for rock friction \cite{34}. It should be noted that $t_{0}\geq\tau$, because the assumption of Eq.\ (1) that we began with is not correct for $t<\tau$ (frictional strength is finite and positive). In any case, at $t_{0}\rightarrow0$ the integral in Eq.\ (3) is divergent and infinite forces are obtained. Such a problem arises in describing the dynamics of all disordered systems demonstrating logarithmic aging \cite{33, 35, 36}. In these systems generally, the logarithmic behavior is valid only after a microscopic time scale of the system.

In the most general form, non-steady state friction under a constant shearing may be a function of both strain or displacement and the shearing velocity. However, dividing Eq.\ (2) by $v$ and making the replacement $x=vt$ denotes the granular friction in terms of sliding distance only. This is, in fact, imposed by our assumption that shear-induced rejuvenation is proportional to the shear rate. We inspect this property of our granular system in the experiments below. However, the rejuvenation, in principle, can be a more complex function, $g(v)$, of the flow rate than assumed here. Rewriting Eq.\ (2) for a rejuvenation term of the general form $\alpha g(v) (F-F_{d})$ shows no change in the functional form of the transient stress overshoot. The linear rejuvenation we use, or equivalently the strain-rate independent transient friction, can be considered as the first-order term in an expansion; we limit ourselves here to this term.

\begin{figure}
	\begin{center}
		\includegraphics[scale=0.6]{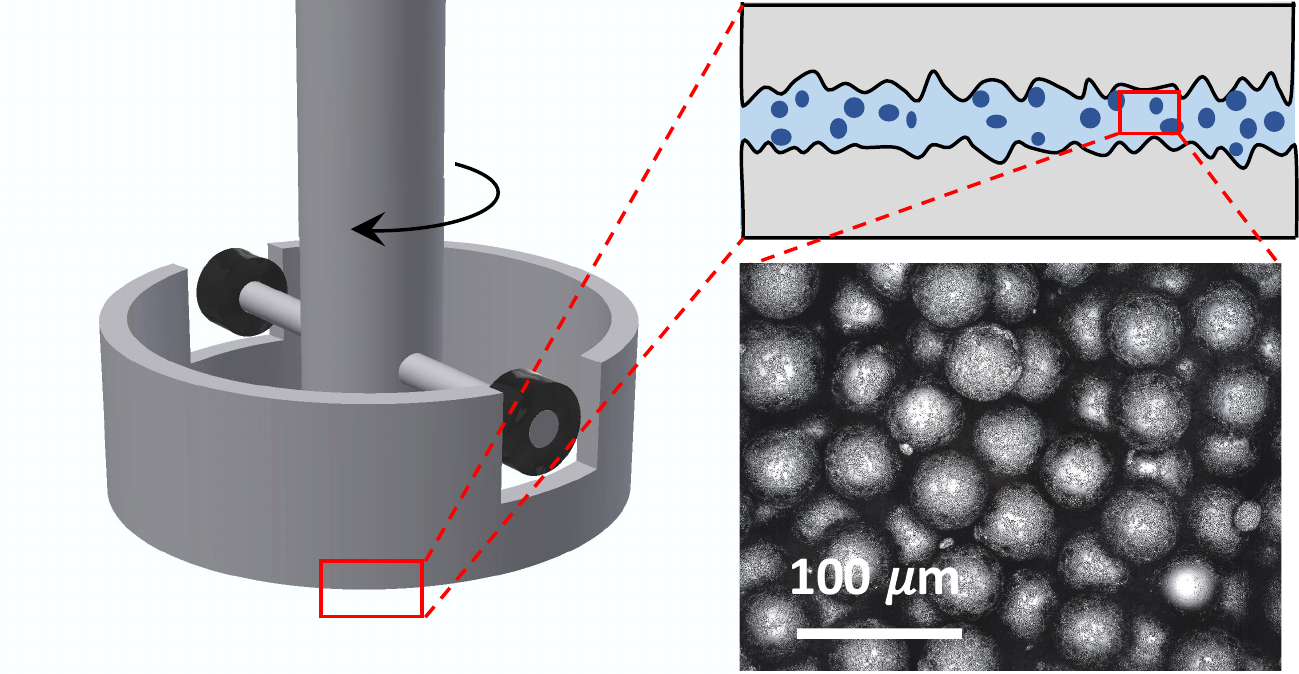}
	\end{center}
	\vspace{-0.5cm}
	\caption{Schematic of the experimental apparatus. A rheometer (Anton Paar MCR 302) is used to rotate a small cut of an aluminum cylindrical tube (outer diameter 29.7 mm, wall thickness of 2.3 mm) around its symmetry axis.  The tube is laying on a plane covered by a $\sim1$ mm layer of PMMA spherical particles of $\sim40$ $\mu m$ diameter. Both the tube cross section and the bottom plate surfaces are sandblasted to have a surface roughness of 2 -3 $\mu m$ (measured with a Keyence optical profilometer). The contact between the rheometer tool and the tube is through frictionless bearing so that the tube is free to move in normal direction. In this way, the rheometer doesn’t apply any normal load to the granular media.}
	\label{f:numeric}\end{figure}

\begin{figure*}[hbt]
	\begin{center}
		\includegraphics[scale=0.73]{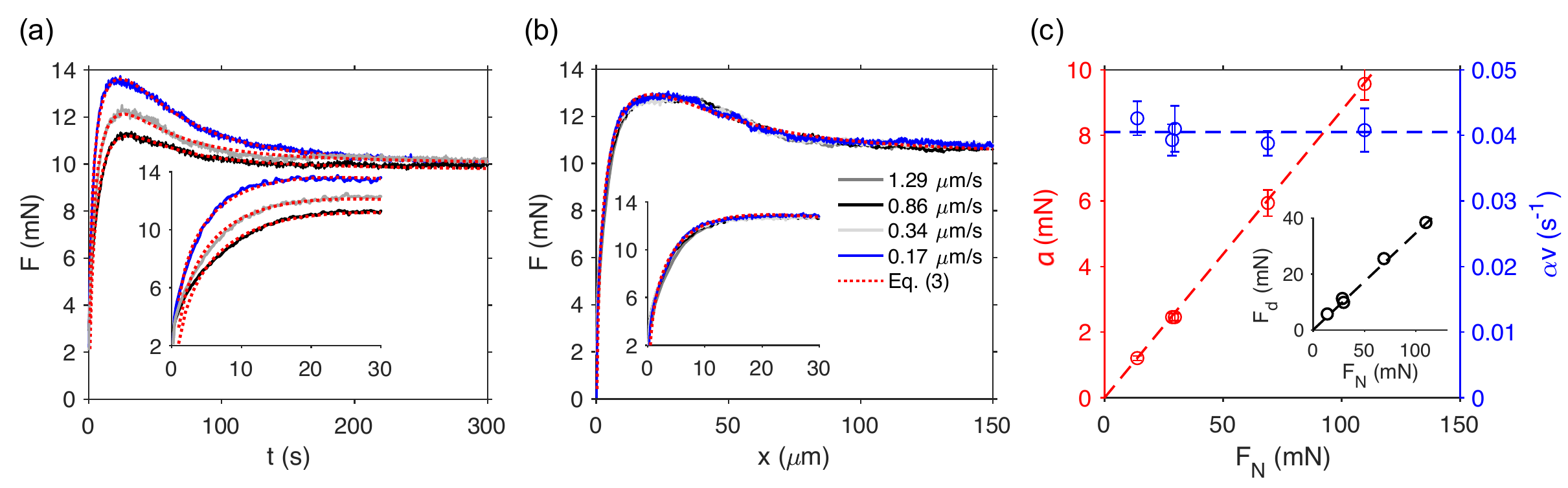}
	\end{center}
	\vspace{-.5cm}
	\caption{Stress overshoot of the onset of granular flow. (a) Friction as a function of time for three typical static to steady flow transitions of the same granular media and system with nominally identical experimental conditions, but prepared at different times. The constant imposed velocity is 0.86 $\mu m/s$. The red dotted lines are the best fits of Eq.\ (3) with $a$, $\alpha v$, and $t_{0}$ respectively as 3.31, 0.036, 0.23 for blue; 2.99, 0.051, 0.39 for gray; and 2.67, 0.058, 0.55 for black curve, with the units of $mN$, $s^{-1}$, $s$. (b) Friction transients as a function of deformation at different sliding velocities for a granular sample that is not changed anymore after the preparation, just rests for 1 min between the tests. The insets show the details of the behaviors at short times. (c) The parameters $a$ (left axis) and $\alpha v$ (right axis) against the total normal load $F_{N}$  (the weight of the tube) on the granular system. The error bars indicate the standard error of at least eight independent measurements. The inset shows the steady-state friction as a function of the normal load. The standard errors in this case are smaller than the marker size.}
	\label{f:numeric}\end{figure*}

Experiments. — Our experimental system is schematically shown in Fig.\ 1. The cross section of a thick cylindrical tube and a plane perpendicular to its axis forms a frictional interface that is separated by a medium of hard spherical polymethyl methacrylate (PMMA) particles. The tube is rotated by a custom-made rheometer tool, while it is free in normal direction. Therefore, independent of the Reynolds dilatancy in the granular material (the volume change with shear deformations) \cite{37, 38, 39, 40}, the normal load on the material is constant and given by the weight of the tube. This is in contrast to granular flows that are constrained by a constant gap, channel, or an opening \cite{15, 17} The other advantage of our experimental setup is using a rheometer that can move a macroscopic object smoothly with high precision and control, while measuring the applied torque. Moreover, the rheometer is able to switch directly between shear rate- and stress-controlled motions, without releasing the existing stress on the material. 

Examples of the measured granular friction as a function of time under a constant imposed rotational rate are presented in Fig.\ 2(a). The different curves are repeats of the same experiment at different times with nominally identical preparations and experimental conditions; and the dotted lines are the best fits of Eq.\ (3). As the graph shows, each of the friction overshoot dynamics are consistent with our aging and rejuvenation description separately. However, while the overshoot of the onset of granular flow always emerges and is in agreement with Eq.\ (3), the peak height and width considerably vary between different preparations of the same system: this is the sensitive dependence on preparation mentioned earlier. Such sensitivity of granular flows to the initial preparation has also been demonstrated in formation of conical granular piles \cite{41}. Importantly, the startup force depends on the preparation, but the steady-state friction does not show this sensitivity. Our formulation and fittings attribute the force variations in the onset of granular flow primarily to changes in $t_{0}$ (see the figure caption). This is a variable that takes a value close to the singular point of integral of Eq.\ (3), i.e.\ $t=0$. Theoretically, $t_{0}$ is a constant of integration that describes the initial state of the granular material before it starts the aging-rejuvenation evolution. 

In Fig.\ 2(b), to avoid the dependence on the preparation, we focus on the friction transients from a single preparation of the tube on the granular layer which is not modified between the consecutive sliding tests. The plot shows the shear resistance of the granular media against the sliding distance at different imposed sliding rates. Before any measurement, we let the system work at rotational rate of $10^{-5}$  $s^{-1}$ (equivalent to sliding velocity of 0.86 $\mu m/s$) for at least one hour to make sure that the granular pile underneath and around the tube does not change during the measurements. Each shear test starts 1 min after the previous shearing was stopped; although the onset stress overshoot of our granular system does not show any significant dependence on this rest time in the range of several minutes (Fig. S1). As evident in the figure, all the friction transients collapse onto a single curve, which is perfectly described by the solution of Eq.\ (3) in terms of the sliding distance. This confirms one of our underlying assumption that the granular mechanics (in the quasi-static regime) is only a function of sliding distance and, notably, does not depend on the shear rate. The steady-state granular friction in the slow shearing regime also does not change with the sliding rate, similarly to the macroscopic friction between solids.

\begin{figure*}[hbt]
	\begin{center}
		\includegraphics[scale=0.8]{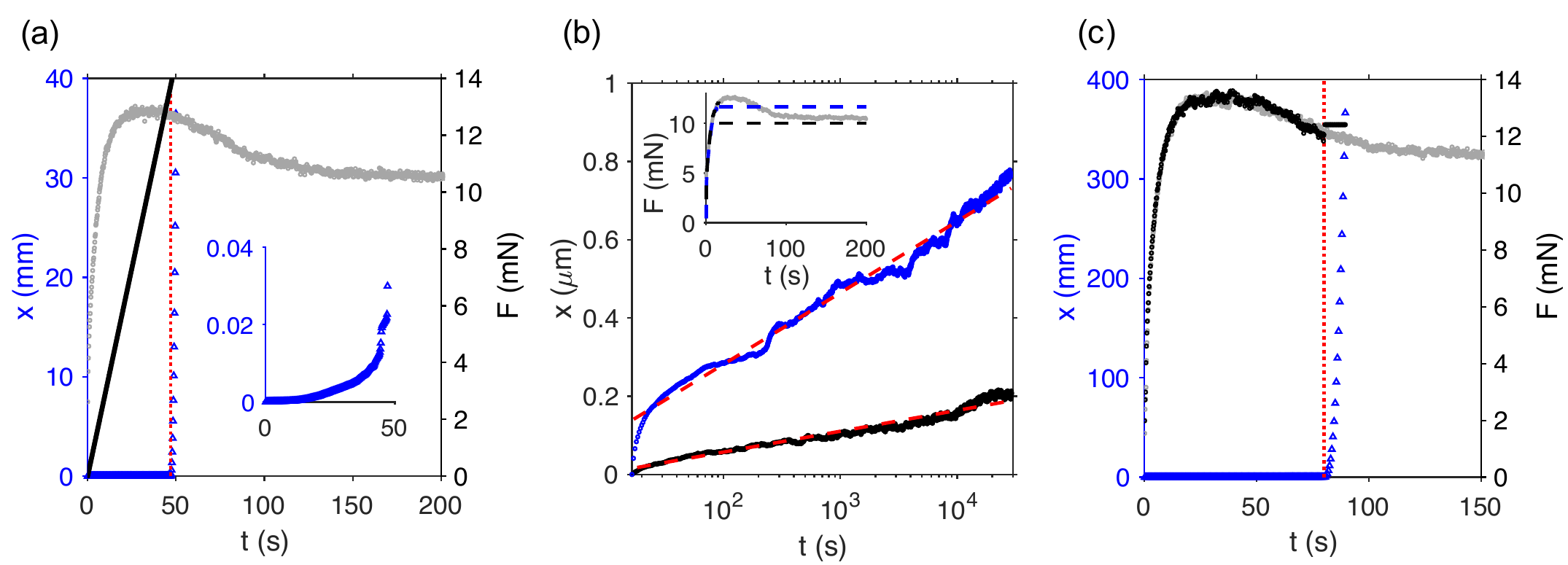}
	\end{center}
	\vspace{-.5cm}
	\caption{Solid-to-liquid failure or unjamming transition of granular matter. (a) A stress ramp (black, right axis) is imposed on a granular sample, which its transient stress overshoot under a constant shear rate (0.86 $\mu m/s$) has been determined before (grey, right axis). The granular material flows suddenly at a point slightly above the peak frictional strength (blue, left axis). The inset magnifies the microscale deformations before the solid-to-liquid failure. (b) Creep deformation of the granular material subjected to constant shear stresses below the threshold friction. The characteristic stress overshoot of the granular sample under a constant shearing (0.86 $\mu m/s$) and the applied shear stresses are shown in the inset. When the applied stress is above the steady-state friction (blue dashed line), the creep deformation is larger. However, the granular structure remains stable and resist the flow. (c) After the peak of the friction overshoot, at time $t = 80$ s, the imposed constant shear rate is instantly switched to a constant shear stress by the rheometer (black, right axis). The constant stress in the second stage is slightly higher than the frictional strength at $t = 80$ s. This immediately leads to the failure and flow of the granular material (blue, left axis). Again, the complete stress overshoot of the sample under a constant shear has been determined before (grey, right axis).}
	\label{f:numeric}\end{figure*}

To further test our first-principle equation for the onset of granular flow, we repeat the shear experiment with five different normal loads on the granular layer. Amontons’ law states that friction is proportional to the normal load on the frictional interface. This law also applies to granular friction \cite{24}. In our slow shear experiments, the steady-state frictional resistance ($F_{d}$) accurately shows this property [Fig. 2(c), inset]. Therefore, Eq. (2) scales correctly over the full range, if $a$ changes proportional to the normal load, and $\alpha$ remains constant. Fig. 2(c) presents $a$ and $\alpha$ as functions of the normal load for multiple friction experiments. Although due to the nonlinearity of friction equations, the overshoot height is very sensitive to the chosen constants, the average results unmistakably confirm that the friction overshoot predicted by Eq. (2) is consistent with the Amontons’ law, and  $a$ and $\alpha$ are in fact the granular system parameters.

We now discuss the consequences of this stress overshoot of the onset of quasi-static granular flows for the (in) stability of stressed granular solids and their transition to flowing liquid-like state. Fig.\ 3(a) shows the generic experiment to measure the threshold (yield) stress: we increase the imposed force (stress) in a stepwise manner (black points) and observe that at a critical stress, all of a sudden, the sample starts to deform rapidly (blue points). If we compare the result of this experiment with the friction force under imposed shear rate, we see that the critical stress corresponds to the maximum of the peak (gray curve). However, as the inset shows, if we zoom in on the displacement curve, there is already significant deformation happening below the threshold stress.  This is a signature of creep under stress that is investigated in Fig.\ 3(b). In this figure, we impose two different force levels, one above and one below the steady-state friction. For both cases, the applied force is smaller than the peak friction; and the peak friction has to be overcome before the system can flow macroscopically. The deformations observed here correspond to a small fraction of a particle diameter (smaller than 1 $\mu m$, compared to the 40 $\mu m$ grain size) over time scales of several hours, so this is not a flow. Nevertheless, it is  interesting to note that both creep curves are logarithmic and the larger force shows a significantly larger creep. This is not unlike our initial assumption that the aging of the system under a quasi-static flow is logarithmic also. The discussion of the static creep however goes beyond the quasi-stationary flow that we describe by Eq. (2) \cite{42}.
	
The conclusion from the above experiments is that the system creeps, but cannot flow for forces smaller than the peak friction. Moreover, experiments imposing a constant strain rate are very different from experiments under an imposed force. This has been observed previously in granular material rheology \cite{38}: there is a large range of shear rates that cannot be attained under an imposed stress, since for an imposed stress the system either flows rapidly, or not at all. In the current work, this is also evident from the experiment in Fig.\ 3(c), where the rheometer first imposes a constant shear rate on the granular sample until the peak frictional resistance has passed and friction is a decreasing function of time. This of course signals an instability, which is triggered if we then (at time $t=80~s$) impose a force that slightly exceeds the one that was measured before: the sample starts to deform rapidly. Conversely, when the same experiment is repeated on the left side of the friction peak, the system stops. This implies that for the same stress the system can either not flow at all, or flow rapidly, showing the importance of the (flow) history of the sample, as predicted also by our simple model. 

In summary, we have presented a comprehensive and consistent picture of the onset of dense granular flows, in which the catastrophic failure of the material is linked to its continuous slow dynamics arising from aging and shear effects. This underlines the fundamental role of aging of the granular packing in the origin of the yield strength. Our simple analytical model accounts for: (1) the rate-independent rheology in the slow-flow regime \cite{18, 28, 29}, (2) creep deformation \cite{42}, (3) the difference between the static and dynamic yield stress \cite{3, 43}, and (4) the sensitive dependence of the flow properties on the preparation of the system \cite{3, 41}; all of which are generic features of granular dynamics. An important implication of the proposed model and the rheology described here is that the mechanical failure of granular fault gouge, leading to an earthquake, could be predicted from analysis of the slow and aseismic slip of the fault and the external stresses.

\vspace*{0.3 cm}
This project has received funding from the European Research Council (ERC) under the European Union’s Horizon 2020 research and innovation program (Grant agreement No. 833240)

\end{document}